\documentstyle[aps]{revtex}
\def\be{\begin{equation}}
\def\ee{\end{equation}}
\def\bi{\bibitem}
\begin{document}
\title{QUANTUM COSMOLOGY WITH CURVATURE SQUARED ACTION}

\draft
\author{{A.K.Sanyal}
\thanks{e-mail: aks@juphys.ernet.in}\\ 
Department of Physics, Jangipur College, Murshidabad, 742213\\ 
and\\ 
{R.C.R.C., department of Physics, Jadavpur University, Calcutta 700032}\\ 
{B. Modak}
\thanks{e-mail: modak@klyuniv.ernet.in}\\
{Department of Physics, Kalyani University, 741235, India}\\
}
\date{\today} 
\maketitle
\begin{abstract}
The correct quantum description for curvature squared term in the action 
can be obtained by casting the action in the canonical form with the 
introduction of a variable which is the negative of the first derivative 
of the field variable appearing in the action, only after removing the 
total derivative terms from the action.We present the Wheeler-deWitt 
equation and obtain the expression for the probability density and 
current density from the equation of continuity, further in the weak 
energy limit we obtain the classical Einstein's equation. Finally we 
present a solution of the wave equation.

\end{abstract}

\section{\bf{Introduction}}

In analogy with the particle quantum dynamics and the quantum field 
theory,Hartle-Hawking \cite {hh:prd} suggested that the ground state wave 
function of the universe is given by

\begin{equation}
\psi_{0}[h_{ij}] = N \int~~\delta g exp(-I_{E}[g]),
\end{equation}

with the proposal that the sum should be taken over all compact four 
geometries. This implies that the universe does not have any boundary in 
the euclidean regime. The above functional integral over all compact four 
geometries bounded by a given three geometry is interpreted as amplitude 
of the given three geometry to arise from a zero three geometry,where 
$\psi_{0}[h_{ij}]$ is finite. Thus the ground state is the amplitude of 
the universe to arise from nothing.
\par
The euclidean form of the Einstein-Hilbert action is not positive 
definite. Therefore the functional integral for the wave function of the 
universe as proposed by Hartle-Hawking runs into serious problems, since 
the wave function diverges badly.There are some prescriptions to make 
the above path integral convergent.In one of the prescriptions \cite 
{jjj:prd}, \cite {jjjb:prd}, \cite {jjl:prd}, it 
was suggested that the proper choice of the contour of integration in the 
functional integral might yield positive definite action, while in 
another prescription \cite {ghp:nupb} a suitable choice of conformal factor 
may give rise to positive definite action. Unfortunately, a completely 
satisfactory result has not yet been obtained through the above two 
prescriptions.

\par
Further to explain the inflationary scenario without invoking phase 
transition in the very early universe, Starobinsky \cite {star:plb} 
considered a field 
equation containing only the geometric terms, but this field equation 
could not be obtained from the action principle,as the terms in the field 
equations are obtained from the perturbative quantum field theory. 
Starobinsky and Schmidt \cite {stm:cqg} later have shown that the 
inflationary phase 
is an attractor in the neighbourhood of the solution of the fourth order 
gravity theory.

\par
However,to get a convergent integral for the wave function Horowitz \cite 
{hor:prd} proposed an action in the form 

\begin{equation} 
S=\frac{1}{4} \int~~d^4 X\sqrt{-g}[A C_{ijkl}^2 +B (R- 4\Lambda)^2],
\end{equation}

where, $C_{ijkl}$ is the Weyl tensor, $R$ is the Ricci scalar,$\Lambda$ 
is the cosmological constant and $A,B$ are coupling constants. In the 
weak energy limit this action (2) reduces to the Einstein-Hilbert action 
for gravity. To obtain a workable and simplified form of the field 
equations, we consider a homogeneous and isotropic minisuperspace model. 
Then $C_{ijkl}$ vanishes trivially and if one chooses $\Lambda =0$, for 
the sake of simplicity, the action (2) retains only the curvature 
squared term.

\par
The classical field equations for $R^2$ gravity can be obtained by the 
standered variational principle. In the variational principle total 
derivative terms in action are extracted and yields a surface integral 
and it is assumed that the surface integral vanishes at boundary of 
variation, or, the action is chosen in such a way that those surface 
integral terms have no contribution in the action. However for 
canonical quantisation this principle is not of much help. For this 
purpose one has to express the action in the canonical form. Boulware 
et al \cite {bou:ah} has proposed that this can be achived by introducing a 
new variable,which is the first derivative of the action with respect to 
the highest derivative of the field variable that appears in the 
action.     

\par
Hawking and Luttrell \cite {hl:nupb} utlised the technique to identify the 
new  
variable and the conformal factor and showed that the Einstein-Hilbert
action along with the curvature squared term reduces to the Einstein's 
gravity being coupled to a massive scalar field. Horowitz \cite {hor:prd} on 
the 
other hand retained only the curvature squared term and showed that the 
corresponding Wheeler-deWitt equation looks very similar to the  
Schr$\ddot{o}$dinger
equation. The semi-classical approximation to such quantum equation has 
also been presented by him. Pollock \cite {pol:nupb}has also introduced 
the same technique to the induced gravity theory and obtained the same 
type of 
result as that obtained by Horowitz \cite {hor:prd} in the sense that the 
corresponding Wheeler-deWitt equation looks similar to the 
Schr$\ddot{o}$dinger equation.

\par
In the present paper we shall deal with the curvature squared term
with a view to study at which stage the new variable should be introduced 
to obtain the correct Wheeler-deWitt equation. Our proposal is that
one should introduce the new variable only after removing all the total 
derivative terms from the action and the new variable is the first 
derivative of the action (free from total derivative terms) with  respect 
to the highest derivative of the field variable. Then one should express
the action in the canonical form with respect to the new variable. At 
this stage one should also remove the total derivative terms (if any) 
from the action and the variation of the action yields the classical 
field equations.

\par
It is important to note that the classical field equations remain the 
same both in our proposal and in the Boulware's \cite {bou:ah} proposal. 
However the canonical quantization in our proposal yields inequivalent 
Wheeler-deWitt equation from that obtained by the technique followed by 
Horowitz \cite {hor:prd}.

\par
To elucidate our proposal we shall consider a toy model of the vacuum 
Einstein-Hilbert action in section 2., and show that the introduction of 
a new variable following Horowitz and Pollock leads to a wrong 
Wheeler-deWitt equation. We therefore conclude that such a variable 
should be introduced only after removing all the total derivative terms 
from the action.

\par
In section 3, we shall consider curvature squared action and with the 
proper choice of the variable ,we show that the corresponding 
Wheeler-deWitt equation is different from that obtained by Horowitz \cite 
{hor:prd}, though it still looks similar to the Schr$\ddot{o}$dinger 
equation.In 
order to compare these results we have chosen FRW model in conformal 
time coordinate for closed ($k=1$) universe.The Wheeler-deWitt 
equation thus obtained contains an effective potential $V$. The 
presence of the potential is in sharp contrast from that obtained by 
Horowitz . The potential $V$ is a function of the expansion 
parameter (i.e. the Hubble parameter)  of the classical field equations 
and it is diverging both at the zero's and at the very large values of 
the expansion parameter. further the effective potential is also 
asymmetric with respect to the expansion and collapse of the universe. 
The physical interpretation of the potential is not  yet clear to us, 
however its extremum yields interesting results. One of the extremum of 
the potential corresponds to the classical vacuum Einstein equation 
,while the other can be viewed upon as a classical equation which 
admits inflation . In fact, the classical field equation obtained from 
the extremum of the potential is not surprising, but it should be a 
desirable feature of higher derivative gravity theory at the weak 
energy limit (i.e. at the long wavelength limit), where the 
contribution of the kinetic energy is sufficiently small with 
comparison to the potential energy in the Hamiltonian; justifying the 
idea of extremum of the potential.

\par
In section 4, we shall present the Hamilton-Jacobi equation 
corresponding to the Schr$\ddot{o}$dinger equation and then we can identify 
the time parameter that appears in the Hamilton-Jacobi equation. In this 
section we shall also present a particular solution of the 
Wheeler-deWitt equation for $n= 0$ operator ordering,which is well 
behaved for all values of the expansion parameters and at the 
vanishing three space volume. The probability density $\psi^*\psi$ is 
a real quantity and is proportional to the scale factor of the universe.

\par
In section 5, we have treated the same problem in proper time 
coordinate for arbitrary three space curvature ($k=o,\pm1$) . This 
shows that the form of the Wheeler-deWitt equation remains unchanged 
. We then turn our attention (see section 6) to compare our formalism to 
that 
developed by Schmidt\cite{sch:prd} earlier. For this purpose we have 
chosen  $k=0$ and have shown that the formalism of Schmidt\cite{sch:prd} 
following 
Ostrogradski [reference cited therein] is different from that of 
ours, though the new variables in both the formalisms turn out to be the 
same and hence the Hamiltonian. This approach of Schmidt has 
recently been considered by Fabris and Reuter \cite{fr:grg}. However in our 
method we have introduced 
the variables in two stages systematically, first, to obtain the field 
equations and then to find the right Hamiltonian that should be 
quantized. The method is simple and straight forward. While the other 
formalism introduced the variables at one stage only, to find the 
Hamiltonian. This formalism would yield the same Hamiltonian as that of 
ours only if one starts from an action free from surface term. 

In section 7, we present a brief conclusion.

\section{\bf{Introducing The New Variable In The Vacuum Einstein-Hilbert 
Action (a toy model)}}

The vacuum Einstein-Hilbert action is

\begin{equation}
S=-\frac{1}{2 \pi^2} \int~~d^4 X \sqrt{-g} R.
\end{equation}

We consider the closed Robertson-Walker metric in the form

\begin{equation}
ds^2 =exp2\alpha(\eta) [d\eta^2 -d\chi^2 -sin^2\chi (d\theta^2 
+sin^2\theta d\phi^2)],
\end{equation}

where $R=-6 exp[-2\alpha](1+\dot{\alpha}^2+\ddot{\alpha})$ and 
$\dot{\alpha}=\frac{d\alpha}{d\eta}$.\\
The action (3) can be expressed as 

\begin{equation}
S= \int~~exp[2\alpha] (1+\dot{\alpha}^2 +\ddot{\alpha}) d\eta 
=\int~~exp[2\alpha] (1-\dot{\alpha}^2) d\eta + S_{m},
\end{equation}

where $S_{m}$ is the surface term given by 
$S_{m}=exp[2\alpha]\dot{\alpha}$.The classical field equation (assuming 
$S_{m}=0$ at the boundary) is 

\begin{equation}
\ddot{\alpha} + \dot{\alpha}^2 + 1 = 0.
\end{equation}

The constraint equation is $H = 0$i.e.,

\begin{equation}
1 + \dot {\alpha}^2 = 0
\end{equation}

and the Wheeler-deWitt equation is 

\begin{equation}
\frac{\partial^2\psi}{\partial\alpha^2} + 
\frac{n}{\alpha}\frac{\partial\psi}{\partial\alpha} -\frac{4}{
\hbar^2}exp[4\alpha] \psi = 0,
\end{equation}

where, $n$ is the operator ordering index. Now instead of removing total 
derivative term from the action (5) if one introduces a new variable $Q$ 
as

\begin{equation}
Q = -\frac{\partial S}{\partial\ddot{\alpha}},
\end{equation}

then we have $Q = - exp [2\alpha]$.Hence the action (5) can be written 
in the canonical form as  

\begin{equation}
S = - \int~~[Q (1+\dot{\alpha}^2) + Q \ddot{\alpha}]d\eta 
=\int~~[\dot{Q}\dot{\alpha}- Q(1+\dot{\alpha}^2)]d\eta - S_{m}.
\end{equation}

one can observe that the surface terms both in (5) and (10) are the 
same. Apart from the surface term in (10) we have

\be
p_{\alpha}=\frac{\partial L}{\partial\dot{\alpha}} = \dot {Q} - 
2Q\dot{\alpha} 
\ee

and

\be
p_{Q} = \frac{\partial L}{\partial \dot {Q}} = \dot{\alpha}.
\ee

Then the classical field equation with respect to the variation of 
$\alpha$ is 

\be
\ddot{Q} - 2Q \ddot{\alpha} -2\dot{Q}\dot{\alpha} = 0.
\ee

One can observe that the above field equation is identically satisfied 
by equation $Q = - exp(2\alpha)$. The other field equation is

\be
\ddot{\alpha} + \dot{\alpha}^2 + 1 = 0.
\ee

Hence,the classical field equations do not change even with the 
introduction of the new variable. We now express the Hamiltonian 
constraint in the phase space variables,so 

\be
H = p_{\alpha}p_{Q} + Q p_{Q}^2 + Q = 0.
\ee

Now canonical quantization should be performed with basic variables 
viz, $\alpha$ and $\dot{\alpha}$ (Boulware et al). Let $\dot{\alpha} = 
x$,so we replace $p_{Q}$ by $x$ and since $Q = -\frac{\partial 
S}{\partial \ddot{\alpha}} = - \frac{\partial L}{\partial\ddot{\alpha}} 
= -\frac{\partial L}{\partial\dot {x}} = - p_{x}$, so $Q$ should be 
replaced by $p_{x}$. Hence 

\be
H = xp_{\alpha} - x^2 p_{x} - p_{x} = 0.
\ee        

Hence the Wheeler-deWitt equation in this case turns out to be

\be
\frac{\partial\psi}{\partial\alpha} = 
\frac{1+x^2}{x}\frac{\partial\psi}{\partial x},
\ee

whose solution is 

\be
\psi = \psi_{0}exp(c\alpha)(1+x^2)^{c/2},
\ee

where $c$ is a constant. Clearly this does not satisfy equation (8). 
Hence the equation (17) is not the correct Wheeler-deWitt equation. 
Therefore we conclude that in order to obtain the correct quantum 
dynamics one should first remove all the total derivative terms from the 
action and then introduce the new variable if at all required. 

\section{\bf{Curvature Squared Action And The Correct Wheeler-deWitt 
Equation}}

We assume the action as 

\be
S = - \frac{m}{4}\int~~(1+\dot{\alpha}^2 +\ddot{\alpha})^2 d\eta,
\ee

where the background geometry is given by metric (4).According to our 
proposal we remove the total derivative terms from the action (19).Hence 

\be
S = - \frac{m}{4}\int~~[(1+\dot{\alpha}^2)^2 +\ddot {\alpha} ^2] d\eta +
 S_{m},
\ee

where $m = 12\pi^2 \beta$ and $S_{m} = - \frac{m}{6} (3 +\dot{\alpha}^2)$ 
.Now we introduce the new variable to cast the action (20) in 
the canonical form

\be
\frac{m}{4} Q = - \frac{\partial S}{\partial\ddot {\alpha}} = m\ddot 
{\alpha}/2. 
\ee

Hence 

\be
S = m/4 \int~~[\dot{\alpha}\dot{Q} - (1 + \dot{\alpha}^2)^2 + Q^2/4] d\eta + 
S_{m_{1}}
\ee

where $S_{m_{1}}$ is the surface term given by 

\be
S_{m_{1}} = - \frac{m}{6}\dot {\alpha} (3 + \dot{\alpha}^2 + 3\ddot{\alpha}).
\ee

Assuming $S_{m_{1}}$ vanishes at the boundary, the classical field 
equations are 

\be
\ddot{Q} - 12\dot{\alpha}^2 \ddot{\alpha} - 4\ddot{\alpha} = 0,
\ee

and 

\be
Q = 2\ddot{\alpha}.
\ee

The definition of $Q$ in (21) is thus recovered in (25) and the equation 
(24) is the correct classical field equation as can be observed by 
replacing $Q$ by $2\ddot{\alpha}$ . Now the Hamiltonian constraint equation 
is 

\be
H = \frac{4}{m} p_{\alpha} p_{Q} + \frac{64}{m^3} p_{Q}^4 + 
\frac{8}{m}p_{Q}^2 - \frac{m}{16} Q^2 + \frac{m}{4},
\ee

where $p_{\alpha}$ and $p_{Q}$ are the canonical momenta corresponding to 
$\alpha$ and $Q$ respectively. Since the canonical quantization should be 
performed with basic variables viz., $\alpha$ and $\dot {\alpha}$, 
therefore choosing $\dot{\alpha} = x$, we replace $p_{Q}$ by $\frac{mx}{4}$ 
and $Q$ by $-\frac{4p_{x}}{m}$ .Hence 

\be
H = xp_{\alpha} - \frac{p_{x}^2}{m} + \frac{m}{4} (1 + x^2)^2 .
\ee

The corresponding Wheeler-deWitt equation is 

\be
i\hbar \frac{\partial\psi}{\partial\alpha} = \frac{\hbar^2}{mx} 
\frac{\partial^2\psi}{\partial x^2} + 
\frac{n\hbar^2}{mx^2}\frac{\partial\psi}{\partial x} + 
\frac{m}{4x}(1+x^2)^2\psi,
\ee

where $n$ is the operator ordering index. Though the Wheeler-deWitt 
equation (28) looks similar to the Schr$\ddot{o}$dinger equation it differs 
from 
that obtained by Horowitz \cite {hor:prd}. For $n=0$, the first derivative 
term in the 
right hand side of (28) disappears, while the Wheeler-deWitt equation 
obtained by Horowitz always retains the first derivative term on the 
right hand side. We have also obtained an effective potential $V(x)$ 
in the process,which has not been obtained by Horowitz. The equation (28) 
can be written as 

\be
i\hbar\frac{\partial\psi}{\partial\alpha} = \hat{H}_ {0}\psi,
\ee

where,

\be
\hat{H}_{0} = \frac{\hbar^2}{mx}(\frac{\partial^2}{\partial x^2} + 
\frac{n}{x}\frac{\partial}{\partial x}) + \frac{m}{4x}(1+x^2)^2.
\ee

It is noted that $\hat{H}_{0}$ operator is  hermitian and as a 
consequence we obtain the equation of continuity 

\be
\frac{\partial\rho}{\partial\alpha} + {\bf{\nabla}}\cdot \bf{J} = 0,
\ee

where $\rho$ and $\bf{J}$ are the probability density and the current 
density respectively for $n = -1$.The $\rho$ and $\bf{J}$ are given by 
$\rho = \psi^*\psi$ and $\bf{J} = (J_{x},0,0)$; where 

\be
J_{x} = \frac{i\hbar}{mx} (\psi^*\psi_{x} - \psi\psi_{x}^*).
\ee

For other operator ordering index $n$ we can also obtain the continuity 
equation, but with respect to a new variable $y$ which is functionally 
related to $x$ only. It is noted that no such continuity equation can be 
obtained from the Wheeler-deWitt equation presented by Horowitz \cite 
{hor:prd}. In 
analogy with the quantum mechanics it is apparent from (29) and (31) 
that the variable $\alpha$ can be identified as the time parameter in 
quantum comosmology, and the variable $x$ acts as a spatial coordinate 
variable in the quantum cosmology. The coordinate variable $x = 
\dot{\alpha} $, which is just the expansion parameter and the Hamiltonian 
operator $\hat{H}_{0}$ diverges at the bounce of the universe i.e., 
$\dot{\alpha} = 0$. Further in the weak energy limit, the contribution 
of the kinetic energy is sufficiently small with respect to the 
potential energy in the Hamiltonian $H_{0}$. At that regime the 
Hamiltonian is almost dominated by the potential energy. Now the 
extremum of the potential $V(x)$ yields 

\be
1 + \dot{\alpha}^2 = 0
\ee

or,

\be
\dot{\alpha}^2 - 1/3 = 0.
\ee

The equation (33) is the classical vacuum Einstein equation. It is to 
be noted that the equation (33) which we obtain here is not 
surprising, rather it should be a desirable feature of higher derivative 
gravity theory at the weak energy limit. Equation (34) on the other 
hand is just the equation admitting inflation, supporting the idea of 
Starobinsky to explain the inflationary scenario without invoking 
phase transition.

\section{\bf{Solution Of The Wheeler-deWitt Equation}}

The Hamilton Jacobi equation can be obtained from (29) and is given by

\be
\frac{\partial A}{\partial\alpha} +\frac{1}{mx} 
(\frac{\partial A}{\partial x})^2 -\frac{m}{4x}(1+x^2 )^2 = 0,
\ee

where $A$ is the Hamilton Jacobi function. This is the exact form of 
the classical Hamiltonian equation given by

\be
\frac{\partial A}{\partial\alpha} +H_{0} = 0,
\ee

where $H_{0}$ is the Hamiltonian and is given by 
$H_{0}=\frac{p_{x}^2}{mx}-\frac{m}{4x}(1+x^2)^2$. The parameter 
$\alpha$ in (36) or (35) is simply the time parameter as we have 
already claimed from equations (28) and (29) in the previous section.

\par
A particular solution of (28) for $n=0$ is given by 

\be
\psi =\psi_{0}exp[\alpha + \frac{im}{2\hbar}(x+\frac{x^3}{3})].
\ee

The wave function $\psi$ is well behaved for all values of $x$ 
(i.e.,for any expansion parameter) and at the vanishing three space 
volume. The probability density $\psi^* \psi$ is just 
$|\psi_{0}|^2exp(2\alpha)$, and it is a real quantity and propor
tional to the scale factor of the universe. However for normalization of 
the probability density one needs boundary condition of the universe.
Recntly, Bachmann and Schmidt \cite{bac:prd} have solved the 
Wheeler-deWitt equation for a minimally coupled scalar field in Bianchi 
type-I model and shown that the notion of a cosmological quantum boundary 
is well defined.  The solution (37) is not a semiclassical one, as the terms within the 
exponential of (37) does not satisfy the Hamilton-Jacobi equation.
Therefore it is not clear wheather such boundary can be called upon in 
the present context.
\section{\bf{Curvature Squared Action In Proper Time 
Coordinate }}

In this section our first attempt is to show that the Wheeler-deWitt 
equation remains unchanged even in the proper time coordinate and then we 
compare our formalism with that of Schmidt \cite{sch:prd} formulated 
earlier. For 
this purpose we consider the FRW metric in proper time coordinate given 
by 

\be 
ds^2 = dt^2 - a^2 [\frac{dr^2}{1-kr^2} +  r^2 (d\theta^2 + sin^2\theta 
d\phi^2)].
\ee

Removing the total derivative terms  the action can now be cast in the 
following form

\be
S = -\frac{m}{4}\int~~[a\ddot{a}^2 + \frac{(\dot{a}^2 +k)^2}{a}] dt + S_{1}, 
\ee

Where dot represents derivative with respect to the proper time and  
$S_{1}=-\frac{m}{2}(k\dot{a} +\dot{a}^3 /3)$. Introducing the new variable

\be
Q =-\frac{4}{m}\frac{\partial S}{\partial\ddot{a}} = 2a\ddot{a} 
\ee

and expressing the action in the canonical form we have 

\be
S = \frac{m}{4}\int~~[\dot{a}\dot{Q} - \frac{(\dot{a}^2 +k)^2}{a} 
+\frac{Q^2}{4a}] dt +S_{2},
\ee

where $S_{2} = -\frac{m}{6}\dot{a} (3k +\dot{a}^2 +3a\ddot{a})$. The canonical 
momenta are

\be
p_{a} = \frac{\partial L}{\partial\dot{a}} = \frac{m}{4} [\dot{Q} - 
4\dot{a}(\dot{a}^2 + k)/a]
\ee

and

\be
p_{Q} = \frac{\partial L}{\partial\dot {Q}} = m\dot{a}/4.
\ee

It is now possible to find the classical field equations, one of which 
turns out to be the definition of $Q$  and the other is found to resemble 
with one already obtained in section 3. The Hamiltonian turns out to be 

\be 
H = \frac{4}{m}(p_{a}p_{Q} + \frac{16p_{Q}^4}{m^2a} + \frac{2kp_{Q}^2}{a} 
+\frac{m^2k^2}{16a} - \frac{m^2 Q^2}{64a}).
\ee

Now, for canonical quantization we invoke the basic variables, i.e., 
choose $\dot{a} = x$, replace $Q$ by $-4p_{y}/m$ and $p_{Q}$ by $my/4$.
Thus the Hamiltonian turns out to be 

\be
H = yp_{y} - \frac{p_{y}^2}{ma} +\frac{m}{4a} (y^2+k)^2, 
\ee

whose quantum version viz., the Wheeler-deWitt equation is 

\be
i\hbar a\frac{\partial\psi}{\partial a} 
=\frac{\hbar^2}{my}(\frac{\partial^2\psi}{\partial y^2} 
+\frac{n}{y}\frac{\partial\psi}{\partial y}) + \frac{m}{4y}(y^2 + k)^2,
\ee

where $n$ is the operator ordering index. Now choosing $a = e^q$ the 
Wheeler-deWitt equation can be written as

\be
i\hbar\frac{\partial\psi}{\partial q} = 
\frac{\hbar^2}{my}(\frac{\partial^2\psi}{\partial y^2} 
+\frac{n}{y}\frac{\partial\psi}{\partial y}) + \frac{m}{4y}(y^2+k)^2.
\ee  

It is interesting to note that this form of Wheeler-deWitt equation is the 
same as that obtained in equation (28) of section 3, in conformal time.

\section {\bf{Comparison With Schmidt's Formalism}}

Let us turn our attention to compare our formalism with that proposed by
Schmidt \cite{sch:prd} earlier. For this purpose 
we choose $k=0$ and $a=e^q$.Thus 
the action (39) turns out to be

\be
S = - \frac{m}{4}\int~~\ddot{q}^2 e^{3q} dt.
\ee

Following the method already proposed, our new variable is 

\be
Q = 2\ddot{q} e^{3q}
\ee

and the canonical action is 

\be
S = \frac{m}{4}\int~~[\dot{q}\dot{Q} + \frac{Q^2}{4}e^{-3q}] dt + S_{3},
\ee

vwhere $S_{3}=-\frac{m}{4}Q\dot{q}$. The canonical momenta are $p_{q} = 
m\dot{Q}/4$ and $p_{Q} = m\dot{q}/4$. Hence the Hamiltonian is 

\be
H = \frac{4}{m}p_{q} p_{Q} - \frac{m}{16} Q^2 e^{-3q}.
\ee

Introducing basic variable $x = \dot{q}$ and replacing $Q$ by $-4p_{x}/m$ 
and $p_{Q}$ by $mx/4$, we find

\be
H = xp_{q} -\frac{p_{x}^2}{m} e^{-3q}.
\ee

Now replacing $q$ by $Q_1$, $\dot{q} = x$ by $Q_2$, $p_{q}=m\dot{Q}/4 = 
\frac{m}{4}\frac{d}{dt}(2\ddot{q} e^{3q})$ by $-mP_1/4$ and finally 
$p_{x}=-mQ/4=-\frac{m}{4}(2\ddot{q} e^{3q})$ by $-mP_2/4$ we can express 
equation (52) as 

\be
H = -\frac{m}{4} (P_1Q_2 + \frac{1}{4}P_2^2 e^{-3Q_1}).
\ee

This Hamiltonian other than the factor $-m/4$ was obtained by Schmidt
\cite{sch:prd} 
following Ostrogradski's prescription. Though the new variables 
introduced in both the cases to construct the Hamiltonian are found to 
match, yet the formalisms differ. In our method after eliminating surface 
terms from the action we have introduced 
variables at two stages, first to cast the action in the canonical 
form for obtaining the classical field equations 
and then to construct the Hamiltonian suitable for quantization, basic 
variables have been called upon. On the other hand the basic 
variables in the Schmidt's proposal have been introduced from the very 
begining and there is no prescription suggested to eliminate the surface 
term. Here lies the difference between the two proposals.The Hamiltonian 
(53) turns out to be the same since Schmidt has considered the action 
(48) which is already free from the surface term. It is not difficult to 
show that Schmidt's formalism would yield a different Hamiltonian if 
action (19) would have been considered.

\section{\bf{Conclusion}}

We conclude that whenever a Lagrangian contains terms higher 
than the first derivative, the correct quantum dynamics would be 
obtained if one casts the action in the canonical form by 
introducing a new variable only after removing all the total 
derivative terms from the action. We have shown that in our 
proposal there is no change in the classical field equations, 
but the quantum dynamics changes sharply as it is evident from 
(29) in comparison with the same equation obtained by Horowitz.            
Therefor, the quantum dynamics is greatly influenced by the presence of 
the surface term in the action. Further the equation of continuity (31) 
identifies the nature of space and time like variables in the 
minisuperspace and therefore establishes the idea of probality density 
and current density in quantum cosmology. In the weak energy limit we are 
able to recover the classical Einstein's field equations from the quantum 
dynamics of higher derivative gravity theory. The solution (37) 
corresponds to a real probability density and it is not a semiclassical 
solution, however for normalization one needs boundary conditions of the 
universe.

\vskip 0.5in

\Large{\bf{Acknowledgement}}\\
The authors like to thank Dr. S.Biswas for valuable 
suggestions.


\begin{references}


\bi{hh:prd} J.B.Hartle and S.Hawking, Phys. Rev. D28 2960 (1993)
\bi{jjj:prd} J.J.Halliwell and J.Louko, Phys. rev. D39 2206 (1989)
\bi{jjjb:prd} J.J.Halliwell and J.B.Hartle, Phys. rev. D41 1815 (1990)
\bi{jjl:prd} J.J.Halliwell and J.Louko, Phys. rev. D42 3997 (1990)
\bi{ghp:nupb} G.W.Gibbons, S.Hawking and M.J.Perry, Nucl.Phys.B138 141 
 (1978) 
\bi{star:plb} A.A.Starobinsky, Phys.Lett.91B 99 (1980)
\bi{stm:cqg} A.A.Starobinsky and H.-J.Schmidt, Class.Quant.Grav.4 695 
 (1987) 
\bi{hor:prd} Gary T.Horowitz, Phys.Rev.D31 1169 (1985)
\bi{bou:ah} D.Boulware,A.Strominger and E.T.Tomboulis in Quantum 
 Theory Of Gravity, edited by S.Christensen (Adam Hilger, Bristol,1984)
\bi{hl:nupb} S.Hawking and J.C.Luttrell, Nucl.Phys.B247 250 (1984)
\bi{pol:nupb} M.D.Pollock, Nucl.Phys.B306 931 (1988)
\bi{sch:prd} H.-J.Schmidt, Phys.Rev.D49 6354 (1994)
\bi{fr:grg} J.C.Fabris and S.Reuter,Gen.Rel.Grav.32 1345 (2000)
\bi{bac:prd} M.Bachmann and H.-J.Schmidt, Phys.Rev.D62 043515 (2000)     

\end{references}
\end{document}